\documentstyle[preprint]{jpsj}

\def\runtitle{Toward Generalized Entropy Composition \\ with Different $q$ Indices 
and 
H-Theorem}
\def\runauthor{Kenichi {\sc Sasaki} and Masahiro {\sc Hotta}}

\title
{
Toward Generalized Entropy Composition \\ with Different $q$ Indices 
and H-Theorem
}

\author
{
Kenichi {\sc Sasaki}\footnote{e-mail:sasaki@tuhep.phys.tohoku.ac.jp}
 and Masahiro {\sc Hotta}\footnote{e-mail:hotta@tuhep.phys.tohoku.ac.jp}
}
\inst
{
Department of Physics, Tohoku University,Sendai 980-8578
}

\recdate
{
\today
}

\abst
{
An attempt is made to construct
composable  composite entropy with different $q$ indices of 
subsystems and address the H-theorem problem of the composite system.
Though the H-theorem does not hold in general situations,
it is shown that some composite entropies do not decrease in time
in near-equilibrium states and factorized states
with negligibly weak interaction between the subsystems.
}

\kword
{
Tsallis statistics, H-theorem, composability, bi-composability
}

\begin{document}
\sloppy
\maketitle

\section{
Introduction
}
For the recent decade, generalization of the Boltzmann-Gibbs 
statistical mechanics has gradually attracted 
much attention. Using the $q$-modified
 entropy, so called Tsallis entropy\cite{t}:
\begin{eqnarray}
S_q =-\frac{1}{1-q}
\left[
1-\sum_i (p_i )^q
\right],
\end{eqnarray}
many anomalous phenomena, including multifractals, self-gravitating systems,
 pure electron plasma respectively, may be described \cite{rv,hp} 
 in "equilibrium" physics language. 
 The index $q$ is a parameter unknown
 a priori and widely believed to be fixed by dynamical details beyond 
 thermodynamical feature of the systems. 
 Here the Tsallis probability variables $p_i$ 
 are related with the physical probability $P_i$ as follows.
\begin{eqnarray}
P_i =\frac{(p_i )^q}{\sum_k (p_k )^q }.
\end{eqnarray}
The entropy is also expressed in terms of the physical variables such that
\begin{eqnarray}
S_q =-\frac{1}{1-q}
\left[
1-\left(\sum_i (P_i)^{\frac{1}{q} } \right)^{-q}
\right] .
\end{eqnarray}
Expectation value of energy $E$ is just given 
in the standard form \cite{tmp} as
\begin{eqnarray}
\left< E \right>=\sum_i E_i P_i .
\end{eqnarray}
Though recently  a subtlety is pointed out  when $P_i$ is treated as the
fundamental variable \cite{abe2},   we in this paper focus on
formulation by use of  the variable $P_i$. 
It is also possible to discuss a composite system of two subsystems $A$ 
and $B$ with the same $q$ indices. The entropy is just extended like 
\begin{eqnarray}
S_{A+B} =-\frac{1}{1-q}
\left[
1-\left(\sum_i \sum_j
(P_{ij})^{\frac{1}{q} } \right)^{-q} 
\right],\label{39}
\end{eqnarray}
and shows the following non-extensivity  for
factorized probability distribution $P_{ij}=P^A_i P^B_j$, 
where $P^A_i$($P^B_j$) is probability for the subsystem $A$($B$).
\begin{eqnarray}
S_{A+B} =S_A +S_B +(1-q)S_A S_B . \label{3}
\end{eqnarray}
Based on the modified entropy one can construct the canonical statistical 
theory and compare its predictions with experimental data of some exotic 
systems. It is known, for example \cite{plasma}, 
that peculiar distribution in 
the quasi-equilibrium state of the electron plasma is well reproduced 
by the theory with $q \sim 0.5$.

Although the generalized statistical theory seems to succeeded in 
 phenomenological descriptions of many exotic systems in which the 
standard statistics does not work, 
we must say that their theoretical foundation has not been fully understood.
For example, it is still unclear whether the second law of the 
thermodynamics of composite systems with {\it different} $q$ subindices holds
 or not. 
Moreover it is impossible so far to discuss  composite entropy 
 of a Tsallis system with $q\neq 1$ and a 
  thermometer made of ordinary matter ($q=1$). 
The essential reason of the poor status  comes from a fact that 
guiding principles for determination of the composite entropy
 have not yet been established.
 
 In this paper, as one of possibilities, 
 we investigate  {\it composable} composite entropy  
  of two Tsallis subsystems $A$ and $B$ 
  with different indices $q_A $ and $q_B$. 
  It is shown that  bi-composability, especially, may possess a powerful
  ability for restriction of the entropy form. 
Here the bi-composability means 
that two kinds of composability are simultaneously
 satisfied. One of the composabilities is just an ordinary one \cite{rv,c,hj}:
\begin{eqnarray}
S_{A+B} (P_{ij} =P^A_i P^B_j )=\lambda (S_A ,S_B). \label{1}
\end{eqnarray}
Thus the total entropy for a factorized distribution 
$P_{ij}=P^A_i P^B_j$ must be computed using only the values of 
 subentropies $S_A$ and $S_B$. By virtue of the composability
  we are able to take a variation of the composite entropy
 directly using the subsystem variations as
\begin{eqnarray}
\delta S_{A+B} =
\frac{\partial \lambda}{\partial S_A} \delta S_A
+
\frac{\partial \lambda}{\partial S_B} \delta S_B .
\end{eqnarray}
Hence it is automatically ensured that the variation of the composite
 entropy vanishes:
\begin{eqnarray}
\delta S_{A+B} =0
\end{eqnarray}
under a deviation as
\begin{eqnarray}
\delta P_{ij} =\delta (P^A_i P^B_j)
=\bar{P}^B_j \delta P^A_i
+\bar{P}^A_i \delta P^B_j ,
\end{eqnarray}
where the distribution $\bar{P}^A_i (\bar{P}^B_j )$ gives extremum of the
 subentropy $S_A (S_B) $. This property will be helpful in building 
 concept of thermal equilibrium even in the extended theory. 
 Here the important point is that even if subsystems are statistically
 independent, there is no guarantee that the composablity should hold. 
 In fact the counter examples can be seen in \cite{a,ap,curado}. 
Another composability is introduced when one constructs a grand 
composite system $(A+B)+(A+B)'$ of two composite systems
 $(A+B)$ and $(A+B)'$. Let us impose the following property on their entropies.
\begin{eqnarray}
S_{(A+B)+(A+B)'} =\Lambda (S_{(A+B)}, S_{(A+B)'}) ,\label{2}
\end{eqnarray}
where $\Lambda$ is an arbitrary function and no need to coincide with 
 the functional form of $\lambda$. This implies that 
the value of 
the grand entropy is fixed only by information of the composite entropies
 $S_{(A+B)}$ and $S_{(A+B)'}$.
The bi-composability is defined by {\it simultaneous} 
realization of the above two 
composabilities (\ref{1}) and (\ref{2}).  
Recall here that 
the Tsallis entropy  satisfies the bi-composability when 
the $q$ indices of the subsystems are the same. Actually 
the first composability holds, just seen in eqn(\ref{3}).
Moreover  a simple computation shows that the second composablity
 is also satisfied as follows.
\begin{eqnarray}
S_{(A+B)+(A+B)'}=S_{(A+B)} +S_{(A+B)'} +(1-q) S_{(A+B)}S_{(A+B)'} .\label{4}
\end{eqnarray}
Meanwhile it may be instructive to note that some generalized entropies  
satisfy one of the composabilities (\ref{1}) and (\ref{2}) but the other not,
 as seen later.
 Therefore the bi-composableness is not automatically induced,
   even assuming a single composability (\ref{1}) or (\ref{2}) and
    statistical independence of the subsystems.

In Appendix A, we give the most general form of the bi-composable
 composite entropy. In later discussions,
  in order to argue as clearly as possible,
  we often explain by use of an attractive toy model, as an example,
  called Tsallis-type bi-composable entropy 
  which satisfies a simple pseudo-additivity relation:
\begin{eqnarray}
S_{(A+B)+(A+B)'}=S_{(A+B)} +S_{(A+B)'} +(1-Q) S_{(A+B)}S_{(A+B)'} ,\label{5}
\end{eqnarray}
even when the subindices take different values: $q_A \neq q_B$.
The property  (\ref{5}) is apparently a straightforward extension 
 of the result (\ref{4}) in the same $q$ index case. 
 The parameter $Q$ may be regarded as grand $q$ index of the 
 composite systems $(A+B)$ and $(A+B)'$. 
 It will be often noticed in later sections
 that a lot of properties which a certain Tsallis-type model has 
 remain still valid for more extended bi-composable entropies.

In this paper 
we also address the H-theorem problem. Though 
several works on the H-theorem for 
 Tsallis statistics have been already performed \cite{m,rs,s},  
 the analysis for composite systems with different q indices
  has not yet been discussed anywhere. 
It is proven in Section 3 
that there exist some probability distributions in which 
 time-derivative of {\it any composable entropy} which satisfies 
  eqn (\ref{1}) takes negative value.
 Meanwhile, it is also analytically shown that the H-theorem for any state
 near equilibrium exactly holds for some 
 bi-composable entropies. Moreover, 
for factorized probability distributions
 with interaction between subsystems negligibly weak, 
the entropies do not decrease in time.

Noncomposable forms of the composite entropy  
may be also the other interesting alternative,
but it looks fairy difficult so far to deal with the entropy systematically,
just in a technical sense. It is beyond scope of this paper 
and we expect progress of the investigation in the near future.

\section{
Composability and Composite Entropy
}
Let us start to write down the most general entropy form for a composite
 system $A+B$ of subsystems $A$ and $B$.
 The probability is denoted by $P_{ij}$ and the subscript $i$ and $j$ 
  are associated with states of $A$ and $B$. The number of the states
 of $A$($B$) is denoted by $N(M)$ in the following.

Note here that for the subsystem $A$($B$) 
we must respect permutation symmetry in the subscripts $i$($j$) 
for the system $A$($B$).
 Even if the numbering of the $A$($B$) states
 are rearranged in an arbitrary way,  the entropy must 
 reproduce the same value as before.
Hence the general form reads
\begin{eqnarray}
S_{A+B}=S_0 (R^A_1 ,R^B_1 ,R^A_2 ,R^B_2,R^A_3 ,R^B_3 ,\cdots ) \label{7}
\end{eqnarray}
where 
\begin{eqnarray}
&&
R^A_\mu =\sum^M_{j=1} \phi^A_\mu 
\left( 
\sum^N_{i=1} \psi^A_\mu (P_{ij})
\right),
\\
&&
R^B_\mu =\sum^N_{i=1} \phi^B_\mu 
\left( 
\sum^M_{j=1} \psi^B_\mu (P_{ij})
\right),
\end{eqnarray}
and $\phi^A_\mu$, $\psi^A_\mu$, $\phi^B_\mu$ and $\psi^B_\mu$
are arbitrary functions. 
Here $\mu $ is a label of function.

Next we construct the most general  composable 
 entropy which is reduced into a function of 
 two subsystem Tsallis entropies $S_{A}$ and $S_{B}$ defined as
\begin{eqnarray}
&&
S_A = 
-\frac{1}{1-q_A }
\left[
1-\left(\sum^N_{i=1} (P^A_i)^{\frac{1}{q_A} } \right)^{-q_A}
\right],
\\
&&
S_B = 
-\frac{1}{1-q_B }
\left[
1-\left(\sum^M_{j=1} (P^B_j)^{\frac{1}{q_B} } \right)^{-q_B}
\right].
\end{eqnarray}
Here $q_A$($q_B$) denotes the $q$ index of the subsystem $A$($B$). 
In order to construct the entropy from eqn (\ref{7}),
it is worthwhile to point out that Tsallis entropies of the subsystems 
have  probability dependence only through
\begin{eqnarray}
&&
C_A = \sum^N_{i=1} (P^A_i )^{r_A} =
\left[
1+(1-q_A ) S_A 
\right]^{-\frac{1}{q_A} },\\
&&
C_B = \sum^M_{j=1} (P^B_j )^{r_B}
=
\left[
1+(1-q_B ) S_B 
\right]^{-\frac{1}{q_B} } ,
\end{eqnarray}
where
\begin{eqnarray}
&&
r_A =\frac{1}{q_A} ,
\\
&&
r_B =\frac{1}{q_B}.
\end{eqnarray}
Taking account of this fact it is easily noticed that 
the first composability (\ref{1}) requires that 
\begin{eqnarray}
S_{A+B} (P^A_i P^B_j )
=S_1 
\left( 
C_A ,C_B , 
\sum^N_{i=1} P^A_i ,
\sum^M_{j=1} P^B_j 
\right) , \label{8}
\end{eqnarray}
where $S_1$ is some function and 
$\sum P^A_i$ and $\sum P^B_j$ will be one after substitution of 
 physical probabilities because of unitarity. 
The relation (\ref{8}) gives a rather strong restriction to 
the functions $R^A_\mu$ and $R^B_\mu$. Let us explain the situation
 using the $R^A_\mu$ case. Substitution of $P_{ij}=P^A_i P^B_j$ into
 $R^A_\mu$ yields
\begin{eqnarray}
\sum^M_{j=1} \phi^A_\mu 
\left( 
\sum^N_{i=1} \psi^A_\mu \left( P^A_i P^B_j \right)
\right)
=
R^A_\mu \left( 
C_A , C_B ,
\sum^N_{i=1} P^A_i ,
\sum^M_{j=1} P^B_j 
\right) .\label{9}
\end{eqnarray}
Because there appears 
 a sum over the subscript $j$ on the left-hand-side
 edge of the left-hand-side term in eqn (\ref{9}),
 $R^A_\mu$ must be always a linear superposition of 
\begin{eqnarray}
\sum^M_{j=1} (P^B_j)^{r_B} f^A_{1\mu}
\left( 
\sum^N_{i=1} (P^A_i )^{r_A},
\sum^N_{i=1} P^A_i ,
\right)\label{10}
\end{eqnarray}
and
\begin{eqnarray}
\sum^M_{j=1} P^B_j f^A_{2\mu}
\left( 
\sum^N_{i=1} (P^A_i )^{r_A},
\sum^N_{i=1} P^A_i ,
\right) .\label{11}
\end{eqnarray}
Here functions $f^A_{1\mu}$ and $f^A_{2\mu} $ are not determined yet.
It should be here stressed again 
that the partial probabilities $P^A_i$
 and $P^B_j$ must enter into the expressions (\ref{10}) and (\ref{11})
 as a simple product $P^A_i P^B_j$. Therefore the form of $R^A_\mu$ is 
 severely constrained and we argue that there exist essentially only 
 the following four candidates to take. From eqn (\ref{10}) 
 the following two components come out:
\begin{eqnarray}
&&
\sum^M_{j=1} (P^B_j)^{r_B} 
\left(
\sum^N_{i=1} (P^A_i)^{r_A}
\right)^{\frac{r_B}{r_A}}
\nonumber\\
&=&
\sum^M_{j=1} 
\left(
\sum^N_{i=1}(P^A_i P^B_j)^{r_A}
\right)^{\frac{r_B}{r_A}}
\nonumber\\
&=&
\sum^M_{j=1}
\left(
\sum^N_{i=1}(P_{ij})^{r_A}
\right)^{\frac{r_B}{r_A}}
\equiv X_1,
\end{eqnarray}

\begin{eqnarray}
&&
\sum^M_{j=1} (P^B_j)^{r_B} 
\left(
\sum^N_{i=1} P^A_i
\right)^{r_B}
\nonumber\\
&=&
\sum^M_{j=1} 
\left(
\sum^N_{i=1} P^A_i P^B_j
\right)^{r_B}
\nonumber\\
&=&
\sum^M_{j=1}
\left(
\sum^N_{i=1} P_{ij}
\right)^{ r_B }
\equiv X_2.
\end{eqnarray}
In the same way the following components can be found using eqn (\ref{11}):
\begin{eqnarray}
&&
\sum^M_{j=1} P^B_j
\left(
\sum^N_{i=1} (P^A_i)^{r_A}
\right)^{\frac{1}{r_A}}
\nonumber\\
&=&
\sum^M_{j=1} 
\left(
\sum^N_{i=1}(P^A_i P^B_j)^{r_A}
\right)^{\frac{1}{r_A}}
\nonumber\\
&=&
\sum^M_{j=1}
\left(
\sum^N_{i=1}(P_{ij})^{r_A}
\right)^{\frac{1}{r_A}}
\equiv X_3,
\end{eqnarray}

\begin{eqnarray}
&&
\sum^M_{j=1} P^B_j  
\left(
\sum^N_{i=1} P^A_i
\right)
\nonumber\\
&=&
\sum^M_{j=1}
\sum^N_{i=1} P_{ij}
=1.
\end{eqnarray}
The last one is trivial and we have finally three-type building blocks
 $X_a$ $(a=1,2,3)$. The same argument is also possible for $R^B_\mu$
  and we conclude that the most general form 
  of the composable composite entropy is given as a function $\Omega$
 in terms of six independent building blocks $X_a$ and $\bar{X}_b$
as follows.
\begin{eqnarray}
S_{A+B} =\Omega (X_a ,\bar{X}_b ;r_A ,r_B) , \label{25}
\end{eqnarray}
where
\begin{eqnarray}
&&
X_1=\sum^M_{j=1}
\left(
\sum^N_{i=1}(P_{ij})^{r_A}
\right)^{\frac{r_B}{r_A}} ,
\\
&&
X_2 =
\sum^M_{j=1}
\left(
\sum^N_{i=1} P_{ij}
\right)^{ r_B } ,
\\
&&
X_3 =\sum^M_{j=1}
\left(
\sum^N_{i=1}(P_{ij})^{r_A}
\right)^{\frac{1}{r_A}} ,
\\
&&
\bar{X}_1=\sum^N_{i=1}
\left(
\sum^M_{j=1}(P_{ij})^{r_B}
\right)^{\frac{r_A}{r_B}} ,
\\
&&
\bar{X}_2 =
\sum^N_{i=1}
\left(
\sum^M_{j=1} P_{ij}
\right)^{ r_A } ,
\\
&&
\bar{X}_3 =\sum^N_{i=1}
\left(
\sum^M_{j=1}(P_{ij})^{r_B}
\right)^{\frac{1}{r_B}} .
\end{eqnarray}
Here permutation symmetry between the subsystems $A$ and $B$
 should be requested in $\Omega$, that is,
\begin{eqnarray}
\Omega (\bar{X}_b ,X_a ;r_B ,r_A)
=
\Omega (X_a ,\bar{X}_b ;r_A ,r_B) .
\end{eqnarray}

So far we have imposed only the first composability (\ref{1}) on the entropy.
It is an interesting problem to look for 
the most general bi-composable form which satisfies
 the two composabilities (\ref{1}) and (\ref{2}). 
 The analysis has been achieved in Appendix A and the result is as follows.
\begin{eqnarray}
S_{A+B} =F(\Delta ; r_A ,r_B), \label{13}
\end{eqnarray}
where $F(x ;r_A ,r_B)$ is an arbitrary function and 
\begin{eqnarray}
\Delta =\prod^3_{a=1} (X_a)^{-\nu_a} 
\prod^3_{b=1} (\bar{X}_b)^{-\bar{\nu}_b} >0,
\label{1000}
\end{eqnarray}
and the exponents $\nu_a$ and $\bar{\nu}_b$ are arbitrary constants.
Apparently  a lot of entropies which are composable but not
 bi-composable can be found. 
 Actually it is possible to construct some entropies
with the form (\ref{25})
 which dependence of $X_a$ and $\bar{X}_b$ occurs not only through $\Delta$.
This simple fact stresses again 
that the notion of bi-composability
is completely independent of a single composability. 
Moreover if one assumes the pseudo-additivity (\ref{5}) for the 
entropy (\ref{13}), it is proven that the entropy must take the 
following form.
\begin{eqnarray}
S_{A+B} =-\frac{1-\Delta}{1-Q} .\label{14}
\end{eqnarray}
The proof is given also in Appendix A. 
The form (\ref{14}) is called  Tsallis-type bi-composable in this paper
 because  double composabilities (\ref{1}) and (\ref{2}) are simultaneously
 satisfied and $\Lambda$ in eqn (\ref{2}) shows the Tsallis-type 
 non-extensivity with its index $Q$. 
 The Tsallis-type  entropy is just a toy model, but
as one of the simplest examples it is quite useful to 
draw what can happen in the composing procedure.

Dynamical details beyond the macroscopic features of the system 
 are expected to determine, in principle, the functional form $\Omega$, or
 $F$, or the values of 
the indices $q_A$, $q_B$ (and $Q$ for the Tsallis-type model), 
and the exponents $\nu_a$, $\bar{\nu}_b$ of the bi-composable entropy, 
 just like determination of  
  the index $q$  in the case with $q_A =q_B =q$ .
 So far we do not know the theoretical method to compute them explicitly.
 However,  once one discovers certain composite systems which 
 are described by canonical ensembles based on 
 the entropies, 
 it is evidently possible that 
 they are measured by parameter 
 fitting analysis of $P_{ij}$. 
The canonical ensemble will be given by extremization of 
the action $S$ defined as
\begin{eqnarray}
S&=&S_{A+B}-\alpha \left( \sum^N_{i=1} \sum^M_{j=1} P_{ij} -1 \right)
\nonumber\\
&&
-\beta
\left( \sum^N_{i=1} \sum^M_{j=1} E_{ij} P_{ij} - \left< E\right> \right) ,
\label{15}
\end{eqnarray}
where $\alpha$ and $\beta$ are Lagrange multipliers and generates
 the unitarity condition:
\begin{eqnarray}
\sum^N_{i=1} \sum^M_{j=1} P_{ij} =1 , \label{16}
\end{eqnarray}
and the total energy constraint of canonical ensemble:
\begin{eqnarray}
\sum^N_{i=1} \sum^M_{j=1} E_{ij} P_{ij} = \left< E\right> ,
\label{17}
\end{eqnarray}
where $\left< E\right>$ is expectation value of the total energy
 and $E_{ij}$ is energy of $(ij)$ state. 
For example, for
 the Tsallis-type bi-composable case,
 the equilibrium equation from eqn (\ref{15}) just reads 
\begin{eqnarray}
\frac{\Delta}{Q-1}
\left[
\sum^3_{a=1} \frac{\nu_a}{X_a}\frac{\partial X_a}{\partial P_{ij}}
+
\sum^3_{b=1} 
\frac{\bar{\nu}_b}{\bar{X}_b}\frac{\partial \bar{X}_b}{\partial P_{ij}}
\right]
=\alpha +\beta E_{ij} .\label{18}
\end{eqnarray}
The eqns (\ref{16}),(\ref{17}) and (\ref{18}) generate the probability
 distribution $P_{ij}$ as a function of $q_A$, 
 $q_B$, $Q$, $\nu_a$ and $\bar{\nu}_b$.
 Consequently one can find out the best fit values of the indices  
 and the exponents from the experimental data which might discover 
 in the future.

\section{
H-theorem
}
In this section let us address the H-theorem problem for 
 the composable entropies. Here we assume the positivity of the subindices:
$r_A >0 ,r_B >0$.
Time evolution of the entropy is now calculated by using 
 the master equation:
\begin{eqnarray}
\frac{dP_{ij}}{dt} = \sum^N_{k=1} \sum^M_{l=1} 
(\Gamma_{ij;kl} P_{kl} -\Gamma_{kl;ij} P_{ij} ),\label{2600}
\end{eqnarray}
where $\Gamma_{ij;kl}$ is transition rate from $(kl)$ state to $(ij)$ state,
 thus takes positive value by definition.

Unfortunately the H-theorem does not hold in general situations.
It is noticed after 
rather tedious calculation that 
 there exist some probability configurations, governed 
 by a special master equational dynamics, 
 in which  {\it the whole composable entropy} (\ref{25}) takes 
 negative values of its time derivative as long as $q_A \neq q_B$. 
 Thus the H-theorem is actually
 broken around the configurations. 
 In order to explain how it fails, 
 let us write down time derivative of the composable entropy (\ref{25}),
  assuming the detailed balance situation:
\begin{eqnarray}
\Gamma_{ij;kl} =\Gamma_{kl;ij},\label{31}
\end{eqnarray} 
  and using the master equation (\ref{2600}) as follows. 
\begin{eqnarray}
\frac{dS_{A+B}}{dt} =\sum_{ijkl}
\Gamma_{ij;kl} S_{ij;kl} , \label{30}
\end{eqnarray}
where
\begin{eqnarray}
S_{ij;kl}
&=&
-\frac{1}{2}
(P_{ij} -P_{kl})
\nonumber\\
&&\times
\left[
\sum_a \frac{\partial \Omega}{\partial X_a}
\left( 
\frac{\partial X_a}{\partial P_{ij}}
-
\frac{\partial X_a}{\partial P_{kl}}
\right)
+
\sum_b \frac{\partial \Omega}{\partial \bar{X}_b}
\left( 
\frac{\partial \bar{X}_b}{\partial P_{ij}}
-
\frac{\partial \bar{X}_b}{\partial P_{kl}}
\right)
\right]. \label{26}
\end{eqnarray}
Next let us calculate the $S_{11;22}$ component in eqn (\ref{26})
 for the following configuration in the case with $(N,M)=(4,4)$.
\begin{eqnarray}
\left[
\begin{array}{cccc}
P_{11} & P_{12} & P_{13} & P_{14} \\
P_{21} & P_{22} & P_{23} & P_{24} \\
P_{31} & P_{32} & P_{33} & P_{34} \\
P_{41} & P_{42} & P_{43} & P_{44} 
\end{array}
\right]
=
\left[
\begin{array}{cccc}
x+\epsilon & 0 & y_{13} & y_{14} \\
0 & x-\epsilon & y_{23} & y_{24} \\
y_{31} & y_{32} & y_{33} & y_{34} \\
y_{41} & y_{42} & y_{43} & 1-2x -\sum y_{cd} 
\end{array}
\right].\label{28}
\end{eqnarray}
What is firstly noticed  here is 
that $S_{11;22}$ vanishes if we take exactly $\epsilon =0$. 
As the result,  expansion of $S_{11;22}$ in terms of $\epsilon$
 behaves as
\begin{eqnarray}
S_{11;22} =-\epsilon K +o(\epsilon^2). \label{27}
\end{eqnarray}
Consequently if $K$ does not vanish in eqn (\ref{27}),
one can always take negative value of $S_{11;22}$
 by taking the same sign of the infinitesimal parameter $\epsilon$
 as that of $K$:
\begin{eqnarray}
S_{11;22} \sim -\epsilon K<0.
\end{eqnarray}
Therefore next let us prove the non-vanishingness of $K$.
When we take $\epsilon =0$, there exist twelve independent parameters
 $x$ and $y_{cd}$ in the expression of $P_{ij}$ (\ref{28}).  
Hence the following twelve variables also are functions of the
twelve parameters:
\begin{eqnarray}
&&
X_a =X_a (x,y_{cd})\ \ \ (a=1,2,3),
\\
&&
\bar{X}_b =\bar{X}_b (x,y_{cd} )\ \ \ (b=1,2,3),
\\
&&
Y_a = 
\frac{\partial X_a}{\partial P_{11}}(x,y_{cd})
-
\frac{\partial X_a}{\partial P_{22}}(x,y_{cd})
\ \ \ (a=1,2,3),
\\
&&
\bar{Y}_b =
\frac{\partial \bar{X}_b}{\partial P_{11}}(x,y_{cd})
-
\frac{\partial \bar{X}_b}{\partial P_{22}}(x,y_{cd})\ \ \ (b=1,2,3).
\end{eqnarray}
After a tedious calculation, 
it can be shown that the Jacobian:
\begin{eqnarray}
J=\left| \frac{\partial (X_a ,\bar{X}_b ,Y_a ,\bar{Y}_b )}
{\partial (x,y_{cd} )}
\right|
\end{eqnarray}
does not vanish as long as $q_A \neq q_B$.
Thus we can regard $X_a ,\bar{X}_b ,Y_a$ and $\bar{Y}_b$ as
 free parameters in the expression of $K$. 
 Because the explicit form of $K$ is given as
\begin{eqnarray}
K=\sum_{a'} \frac{\partial \Omega}{\partial X_{a'}}
(X_a ,\bar{X}_b ) Y_{a'}
+
\sum_{b'} \frac{\partial \Omega}{\partial \bar{X}_{b'}}
(X_a ,\bar{X}_b ) \bar{Y}_{b'} ,
\end{eqnarray}
it is trivial that one can always choose the independent parameters
 to achieve $K\neq 0$.
 Therefore we conclude that there exist the configurations
 for which $S_{11;22}$ is negative. 
Consequently if we set transition rates $\Gamma_{ij;kl} $ to zero except
\begin{eqnarray}
\Gamma_{11;22}=\Gamma_{22;11}=1,
\end{eqnarray}
decrease of the entropy for the configuration:
\begin{eqnarray}
\frac{dS_{A+B}}{dt}<0
\end{eqnarray}
is easily shown due to eqn (\ref{30}). 
 Hence we have found breakdown of the H-theorem
 for {\it any} composable entropy (\ref{25}) 
 in the configuration (\ref{28}).

This failure  might give a negative impression for 
 the composable entropy itself, 
but we do not think that the situation is so serious. 
This is because  the q-modified 
 statistical picture itself might be inadequate to describe 
 the physics for configurations like eqn (\ref{28}). 
  There is a possibility that
 the entropy permits the H-theorem to hold  for 
 physically relevant configurations like near-equilibrium states,
  where the q-deformed statistical description is able to work enough.
In fact we can find some Tsallis-type bi-composable
 models as the examples 
 in which the H-theorem rigidly holds around the equilibrium. 
When the detailed balance transition rate (\ref{31}) is adopted,
the equilibrium distribution of the model 
is just equal-weight profile:
\begin{eqnarray}
\bar{P}_{ij}=\frac{1}{NM}.
\end{eqnarray}
Near the equilibrium the probability is expressed as follows.
\begin{eqnarray}
P_{ij} =\bar{P}_{ij} +\epsilon_{ij} ,\label{32}
\end{eqnarray}
where $\epsilon_{ij}$ are infinitesimal variables and satisfy
 the unitary condition:
\begin{eqnarray}
\sum_{ij} \epsilon_{ij}=0 .
\end{eqnarray}
For the configuration (\ref{32}) we are able to compute $S_{ij;kl}$
 of the Tsallis-type bi-composable model (\ref{14}) as follows.
\begin{eqnarray}
S_{ij;kl}&=&
-\frac{NM\Delta}{2(Q-1)}
(\epsilon_{ij} -\epsilon_{kl} )
\nonumber\\
&&
\times
\left[
\{ r_B (r_A -1)\nu_1 +r_A (r_B-1) \bar{\nu}_1 
+(r_A -1) \nu_3 +(r_B -1) \bar{\nu}_3 \}
(\epsilon_{ij} -\epsilon_{kl} )
\right.
\nonumber\\
&&
+\{ r_B (r_B -r_A ) \nu_1 +r_B (r_B -1) \nu_2 -(r_A -1)\nu_3\}
\frac{1}{N}\sum^N_{s=1}(\epsilon_{sj} -\epsilon_{sl} )
\nonumber\\
&&
\left.
+\{ r_A (r_A -r_B ) \bar{\nu}_1 
+r_A (r_A -1) \bar{\nu}_2 -(r_B -1)\bar{\nu}_3 \}
\frac{1}{M}\sum^M_{s=1}(\epsilon_{js} -\epsilon_{ks} )
\right]
\nonumber\\
&&+o(\epsilon^3).\label{35}
\end{eqnarray}
Therefore if one chooses the grand index $Q$ and the exponents 
$\nu_a$, $\bar{\nu}_b$ of the model to allow the relations:
\begin{eqnarray}
&&
g\equiv\frac{
r_B(r_A -1) \nu_1 +r_A (r_B -1) \bar{\nu}_1
+(r_A -1)\nu_3
+(r_B -1)\bar{\nu}_3
}{1-Q} >0, \label{36}
\\
&&
r_B (r_B -r_A ) \nu_1 +r_B (r_B -1) \nu_2 -(r_A -1)\nu_3 =0 ,\label{37}
\\
&&
r_A (r_A -r_B ) \bar{\nu}_1 
+r_A (r_A -1) \bar{\nu}_2 -(r_B -1)\bar{\nu}_3 =0,\label{38}
\end{eqnarray}
all the coefficients $S_{ij;kl}$ take non-negative values as 
\begin{eqnarray}
S_{ij;kl}\sim\frac{1}{2}gNM\Delta (\epsilon_{ij}-\epsilon_{kl} )^2
\geq 0 .
\end{eqnarray}
Consequently it is concluded using both eqn (\ref{30}) and positivity of 
$\Gamma_{ij;kl}$ that 
the value of the entropy  does not decrease in time 
 around the equilibrium:
\begin{eqnarray}
\frac{dS_{A+B}}{dt} \geq 0.\nonumber
\end{eqnarray}
It is also possible to construct a 
simple and regular toy model of the Tsallis-type entropy satisfying 
eqns (\ref{36}),(\ref{37}) and (\ref{38}), 
 and it is given in Appendix II.

Furthermore, beyond the Tsallis-type model, 
let us introduce a more general bi-composable entropy
 as
\begin{eqnarray}
S_{A+B} =G\left( -\frac{1-\Delta}{1-Q} \right)
\label{2001}
\end{eqnarray}
which satisfies
\begin{eqnarray}
\frac{d}{dx}G(x) >0, \label{2002}
\end{eqnarray}
 and 
eqns (\ref{36}), (\ref{37}) and (\ref{38}).
Then it is trivial that the H-theorem around the equilibrium also holds 
for the entropy (\ref{2001}) due to monotonically increasingness of $G$.

It is quite remarkable that 
the  bi-composable entropy (\ref{2001}) 
 does not decrease in time when interaction between the subsystems is 
 negligibly weak 
 and initial probability takes factorized form as
\begin{eqnarray}
P_{ij} (0) =P^A_i (0) P^B_j (0) .\label{19}
\end{eqnarray}
Due to the omission of the interaction, the transition rates can be 
written as
\begin{eqnarray}
\Gamma_{ij;kl} = \Gamma^A_{i;k}\delta_{jl} +
\delta_{ik} \Gamma^B_{j;l} , \label{20}
\end{eqnarray}
where $\Gamma^A_{i;k}$($\Gamma^B_{j;l}$) is transition rate 
 for the subsystem $A$($B$). 
 The factorized initial condition (\ref{19}) and eqn (\ref{20}) 
 enable  the master equation to be reduced
  into two subsystem equations as
\begin{eqnarray}
&&
\frac{dP^A_i}{dt}=\sum^N_{k=1} 
(\Gamma_{i;k}^A P^A_k -\Gamma^A_{k;i} P^A_i ),
\\
&&
\frac{dP^B_j}{dt}=\sum^M_{l=1} 
(\Gamma_{j;l}^B P^B_l -\Gamma^B_{l;j} P^B_j ),
\end{eqnarray}
and guarantees the factorized time-evolution form:
\begin{eqnarray}
P_{ij} (t) =P^A_i (t) P^B_j (t).
\end{eqnarray}
Here it is convenient to introduce a R{\'e}nyi-type entropy as
\begin{eqnarray}
S_R =\frac{1}{1-Q} \ln \Delta. \label{21}
\end{eqnarray}
Then the time-derivative of the entropy (\ref{2001}) is related with
 that of the R{\'e}nyi-type entropy (\ref{21}) as follows.
\begin{eqnarray}
\frac{dS_{A+B}}{dt} =\Delta G'\left(-\frac{1-\Delta}{1-Q} \right) 
\frac{dS_R}{dt}.
\end{eqnarray}
Also it is a useful property that the R{\'e}nyi-type entropy (\ref{21}) 
 has additivity up to factors like that
\begin{eqnarray}
S_R (P^A_i P^B_j ) =c_A S_{R,A}+c_B S_{R,B},
\end{eqnarray}
where $S_{R,r_A}$ and $S_{R,r_B}$ are the standard R{\'e}nyi entropies 
\cite{r}
 defined by
\begin{eqnarray}
&&
S_{R,A}=\frac{1}{1-r_A }\ln \left( 
\sum^N_{i=1} (P_i^A )^{r_A}
\right),
\\
&&
S_{R,B}=\frac{1}{1-r_B}\ln \left( 
\sum^M_{j=1} (P_j^B )^{r_B}
\right),\label{00021}
\end{eqnarray}
and $c_A$ and $c_B$ are constants defined as
\begin{eqnarray}
&&
c_A =\frac{1-q_A}{1-Q} 
\left(
r_B \nu_1 +r_A \bar{\nu}_1 
+r_A \bar{\nu}_2 +\nu_3
\right),\label{002}
\\
&&
c_B =\frac{1-q_B}{1-Q} 
\left(
r_B \nu_1 +r_A \bar{\nu}_1 
+r_B \nu_2 +\bar{\nu}_3
\right).\label{003}
\end{eqnarray}
It has been already known \cite{m,rs} that 
the R{\'e}nyi entropy for positive index monotonically increases 
 by the dynamics of the master equation:
\begin{eqnarray}
&&
\frac{dS_{R,A}}{dt} \geq 0,
\\
&&
\frac{dS_{R,B}}{dt} \geq 0.
\end{eqnarray}
 By using eqns (\ref{37}) and (\ref{38}) in the expressions of
  $c_A$ and $c_B$, it can be shown that
\begin{eqnarray}
&&
c_A>0, \label{22}
\\
&&
c_B >0 \label{23}
\end{eqnarray}
 if eqns (\ref{36})  holds.
 Therefore it turns out that 
 time-derivative of the entropy (\ref{2001}) entropy 
 takes non-negative value:
\begin{eqnarray}
\frac{dS_{A+B}}{dt} \geq 0, \label{24}
\end{eqnarray}
because the inequality:
\begin{eqnarray}
\frac{dS_R}{dt}=c_A \frac{dS_{R,A}}{dt} +c_B \frac{dS_{R,B}}{dt} \geq 0
\end{eqnarray}
holds and
\begin{eqnarray}
\Delta G'\left(-\frac{1-\Delta}{1-Q} \right) >0
\end{eqnarray}
 is guaranteed due to eqn (\ref{2002}).
Consequently it has been proven that there exists  the
 composable entropies which satisfy 
 the H-theorem for both 
 the near-equilibrium case and the uncorrelated subsystems case, and
 an explicite example is the form (\ref{2001}).

It may be interesting to comment, along Abe's argument\cite{abe}
 (also see \cite{rj}), on the thermal balance relation
 generated by the entropy (\ref{2001}), though the discussion
 is still in a rather heuristic level. It turns out that 
the nonextensivity of the entropy (\ref{2001})
is given as follows.
\begin{eqnarray}
S_{A+B} &=& G\left( -\frac{1-\bar{\Delta}}{1-Q} \right), \nonumber\\
\bar{\Delta} &=&
\left[
1+(1-q_A ) S_A 
\right]^{r_B \nu_1 +\nu_3 +r_A \bar{\nu}_1 +r_A \bar{\nu}_2 }
\nonumber\\
&&\times
\left[
1+(1-q_B ) S_B 
\right]^{r_A \bar{\nu}_1 +\bar{\nu}_3 +r_B \nu_1 +r_B \nu_2 } .
\end{eqnarray}
Using the H-theorem conditions (\ref{37}) and (\ref{38}),
the variation of the entropy can be written as
\begin{eqnarray}
\delta S_{A+B}
&=&\frac{r_B (r_B -1) (\nu_1 +\nu_2 )
+r_A (r_A -1) (\bar{\nu}_1 +\bar{\nu}_2 ) }{1-Q}\Delta G' 
\nonumber\\
&&
\times
\left[
\frac{q_A}{1+(1-q_A) S_A}\delta S_A
+
\frac{q_B}{1+(1-q_B) S_B}\delta S_B
\right] .
\end{eqnarray}
Then we take $\delta S_{A+B}=0$
under the total energy conservation relation:
\begin{eqnarray}
\delta E_A +\delta E_B =0
\end{eqnarray}
to get the thermal balance relation.  
The result is given as follows,
 independent of the functional form $G(x)$ and the value of $Q$.
\begin{eqnarray}
\frac{q_A}{1+(1-q_A) S_A}\frac{\delta S_A}{\delta E_A}
=
\frac{q_B}{1+(1-q_B) S_B}\frac{\delta S_B}{\delta E_B} .\label{13000}
\end{eqnarray}
Note that this relation includes the Abe's balance relation \cite{abe} 
as a special case.
Actually when $q_A =q_B =q$ is taken 
\begin{eqnarray}
\frac{1}{1+(1-q) S_A}\frac{\delta S_A}{\delta E_A}
=
\frac{1}{1+(1-q) S_B}\frac{\delta S_B}{\delta E_B}
\end{eqnarray}
is exactly reproduced. Also eqn (\ref{13000}) is 
 consistent with a guessed relation \cite{c} 
 by Tsallis for the different q case.
Though we have a subtlety that 
 the canonical thermal state of the composite noninteracting system
 is not the product of the states of subsystems for the generalized
  entropy, 
 there is a possiblity that the above result (\ref{13000}) remains unchanged 
 for thermodynamic-limit situations just as in the same q case\cite{abe} 
 (also see \cite{c}).

If the system $A$ is taken as an ordinary Boltzmann-Gibbs system
($q_A =1$), the subentropy is reduced into the BG form $S_{BG:A}$. 
Then the relation
(\ref{13000})  is re-expressed as 
\begin{eqnarray}
\frac{\delta S_{BG:A}}{\delta E_A}
=
\frac{q_B}{1+(1-q_B) S_B}\frac{\delta S_B}{\delta E_B}.
\end{eqnarray}
Even in the extotic situation,
 the fact can be considered robust 
 that physical temperature $T_{phys}$ can be 
introduced for the system $A$ as follows.
\begin{eqnarray}
\frac{1}{T_{phys}}=\frac{\delta S_{BG:A}}{\delta E_A}.
\end{eqnarray}
Therefore observable temperature $T_B$ of the Tsallis system $B$ should be
defined as 
\begin{eqnarray}
\frac{1}{T_B}=\frac{q_B}{1+(1-q_B) S_B}\frac{\delta S_B}{\delta E_B},
\label{15000}
\end{eqnarray}
so as to preserve the zero-th law of thermodynaics:
\begin{eqnarray}
\frac{1}{T_{phys}} =\frac{1}{T_B}.
\end{eqnarray}
Here we want to mention that before our analysis 
no one argues explicitly presence of the numerator $q_B$ of the prefactor 
 in the right-hand-side term of eqn (\ref{15000}). 
Due to the definition (\ref{15000}), the
original relation (\ref{13000}),  in which $q_A$ is not needed to take unit,
can be interpreted as a generalized thermal balance as follows.
\begin{eqnarray}
\frac{1}{T_A} =\frac{1}{T_B}.\label{14000}
\end{eqnarray}
The transitivity relation (\ref{14000}) looks quite plausible
 and attractive, though the derivation seems still nonunique.

\ \\

In summary,  though the H-theorem of all the composable entropies (\ref{25})
 does not hold for some 
 probability distributions like in eqn (\ref{28}),
  it has been rigorously proven that 
there exists  a
rather general entropy (\ref{2001}) which does not decrease in time 
 for both any states around the equipartion equilibrium
 and the factorized distributions 
 without interaction between the subsystems. 
We also derive the thermal balance relation for the different-indices case
 from a plausible but still  heuristic argument.  
 
 Though 
 analysis for noncomposable composite entropy has not been included in this
 paper, the possibility remains, of course, 
 still alive and may be interesting. 
 However the systematic investigation seems complicated and
 seems to need some innovation in the future.

\section*{Acknowledgments}
We would like to thank I. Joichi, M. Morikawa,  
 A. K. Rajagopal and C. Tsallis for their valuable comments.
 The work of K. S. is supported by a fellowship
  of the Japan Society of the Promotion of Science.

\appendix
\section{Bi-Composable Entropy}
In this appendix, let us prove first that 
 the most general bi-composable entropy takes
 the following form.
\begin{eqnarray}
S_{A+B} =F(\Delta ;r_A ,r_B ),\nonumber
\end{eqnarray}
where the function $F$ is arbitrary and $\Delta$ is defined in eqn 
(\ref{1000}). Recall here that the most
 general composable entropy is given in eqn (\ref{25}).
 Thus our remaining task is just imposition of 
 another composability (\ref{2}) on $\Omega$ in eqn (\ref{25}).
 To achieve this, it is very helpful to note a fact that
\begin{eqnarray}
&&
\sum^N_{i=1}\sum^{N'}_{i'=1}
\left[
\sum^M_{j=1}\sum^{M'}_{j'=1}
\left( P^{(A+B)}_{ij} P^{(A+B)'}_{i'j'} \right)^\alpha
\right]^\beta
\nonumber\\
&&=
\left( 
\sum^N_{i=1} \left[\sum^M_{i=1} \left( 
P^{(A+B)}_{ij}
\right)^\alpha \right]^\beta
\right)
\left( 
\sum^{N'}_{i'=1} 
\left[\sum^{M'}_{i'=1} \left( 
P^{(A+B)'}_{i'j'}
\right)^\alpha \right]^\beta
\right).
\end{eqnarray}
Thus all the six components $X_a $ and $\bar{X}_b$ are completely
 factorized for the probability forms as
\begin{eqnarray}
P_{ijkl} =P^{(A+B)}_{ij} P^{(A+B)'}_{kl}.
\end{eqnarray}
Consequently we obtain the following parametric relation among 
 $S_{(A+B)}$, $S_{(A+B)'}$ and $S_{(A+B)+(A+B)'}$.
\begin{eqnarray}
&&
S_{(A+B)+(A+B)'} =\Omega
(s_a t_a ,\bar{s}_b \bar{t}_b ; r_A ,r_B),
\label{007}
\\
&&
S_{(A+B)} =\Omega (s_a , \bar{s}_b ; r_A ,r_B),
\label{008}
\\
&&
S_{(A+B)'}=\Omega (t_a , \bar{t}_b ; r_A , r_B),
\label{009}
\end{eqnarray}
where 
\begin{eqnarray}
&&
s_a =X^{(A+B)}_a ,
\\
&&
t_a =X^{(A+B)'}_a ,
\\
&&
\bar{s}_b =\bar{X}^{(A+B)}_b ,
\\
&& 
\bar{t}_b =\bar{X}^{(A+B)'}_b
\end{eqnarray}
 can be regarded as free parameters of the expression, 
 independent from each other by taking $N,N',M$ and $M'$ largely enough. 
 Using the relations (\ref{007})$\sim$(\ref{009}), 
 it is possible to write down a lot of parametric expressions of 
 $S_{(A+B)+(A+B)'}$ in terms of $S_{(A+B)}$ and $S_{(A+B)'}$.
 For example, we obtain an expression by taking
\begin{eqnarray}
&&
s_1 =s_2 =s_3 =\bar{s}_1 =\bar{s}_2 =\bar{s}_3 =\phi,
\label{0010}
\\
&&
t_1 =t_2 =t_3 =\bar{t}_1 =\bar{t}_2 =\bar{t}_3 =\theta,
\label{0011}
\end{eqnarray}
 as follows.
\begin{eqnarray}
&&
S_{(A+B)+(A+B)'} =\Omega
(\phi\theta,\phi\theta,\phi\theta,\phi\theta,\phi\theta,\phi\theta ;
r_A ,r_B),
\label{0012}
\\
&&
S_{(A+B)}=\Omega (\phi,\phi,\phi,\phi,\phi,\phi ; r_A ,r_B),
\label{0013}
\\
&&
S_{(A+B)'}=\Omega (\theta,\theta,\theta,\theta,\theta,\theta ; r_A ,r_B).
\label{0014}
\end{eqnarray}
The above relations (\ref{0012})$\sim$(\ref{0014}) indicate that
 $S_{(A+B)+(A+B)'}$ is given as a function of $S_{(A+B)}$ and 
 $S_{(A+B)'}$ implicitly via the two parameters $\phi$ and $\theta$.

Next we must find explicitly the general solution $\Omega$ 
of eqns (\ref{007})$\sim$(\ref{009}). In order to do that,
 let us consider first the following relation obtained from
 eqns (\ref{008}) and (\ref{0013}).
\begin{eqnarray}
&&
\Omega (s_a ,\bar{s}_1 ,\bar{s}_2 ,\bar{s}_3 ;r_A ,r_B)
\nonumber\\
&&
=
\Omega (\phi,\phi,\phi,\phi,\phi,\phi ; r_A ,r_B).
\label{0015}
\end{eqnarray}
By solving eqn (\ref{0015}) for $\bar{s}_3$ , we can define 
a function $\sigma$ as
\begin{eqnarray}
\bar{s}_3 =\sigma (\phi,s_a ,\bar{s}_1 ,\bar{s}_2).
\label{0016}
\end{eqnarray}
Substituting eqn (\ref{0016}) into eqn (\ref{007}) and 
using eqn (\ref{0012}) yield 
\begin{eqnarray}
&&
\Omega
(s_a t_b ,\bar{s}_1 \bar{t}_1 ,\bar{s}_2 \bar{t}_2 ,
\sigma(\phi ,s_a,\bar{s}_1 ,\bar{s}_2 )
\sigma(\theta ,t_a,\bar{t}_1 ,\bar{t}_2 ) ;
 r_A ,r_B)
 \nonumber\\
 &&
 =
 \Omega
(\phi\theta,\phi\theta,\phi\theta,\phi\theta,\phi\theta,\phi\theta ;
r_A ,r_B).
\label{0017}
\end{eqnarray}
Then it is soon noticed that eqn (\ref{0017}) means 
\begin{eqnarray}
&&
\sigma(\phi ,s_a,\bar{s}_1 ,\bar{s}_2 )
\sigma(\theta ,t_a,\bar{t}_1 ,\bar{t}_2 ) 
\nonumber\\
&&
=Z(s_a t_a ,\bar{s}_1 \bar{t}_1 ,\bar{s}_2 \bar{t}_2 ,\phi\theta),
\label{0018}
\end{eqnarray}
where $Z$ is a certain function. Eqn (\ref{0018}) looks very complicated,
but fortunately can be solved by treating the variables 
as six pairs: $(s_a ,t_a)$,$(\bar{s}_1,\bar{t}_1)$,
$(\bar{s}_2 ,\bar{t}_2)$ and $(\phi,\theta)$. For each pair  we  
 solve the equation independently. Actually for an arbitrary pair
  $(\rho_s ,\rho_t )$ in the six, eqn (\ref{0018}) is regarded as
\begin{eqnarray}
\sigma (\rho_s )\tilde{\sigma}(\rho_t )
=Z(\rho_s \rho_t).
\label{0019}
\end{eqnarray}
>From eqn (\ref{0019}) the following relations are straightforwardly
 obtained by taking $\rho_s =1$ or $\rho_t =1$.
\begin{eqnarray}
&&
\sigma (\rho) =\frac{1}{\tilde{\sigma} (1)}Z(\rho),
\label{0020}
\\
&&
\tilde{\sigma} (\rho) =\frac{1}{\sigma (1)}Z(\rho).
\label{0021}
\end{eqnarray}
Substituting eqns (\ref{0020}) and (\ref{0021}), eqn (\ref{0019})
 is replaced into 
\begin{eqnarray}
Z(\rho_s )Z(\rho_t)=Z(1) Z(\rho_s \rho_t).
\label{0022}
\end{eqnarray}
By taking differentiation with respect to $\rho_t$ in eqn (\ref{0022})
 and setting $\rho_t =1$ and $\rho_s =\rho$, the following equation
 appears.
\begin{eqnarray}
\rho\frac{dZ}{d\rho}
=\frac{Z'(1)}{Z(1)}Z(\rho),
\label{0023}
\end{eqnarray}
where $Z'(1)$ is derivative of $Z$ at $\rho=1$.
This equation is easily integrated and leads us to the following
 relation:
\begin{eqnarray}
\sigma(\rho) = \frac{1}{\tilde{\sigma}(1)} Z(\rho) \propto \rho^\xi ,
\end{eqnarray}
where $\xi$ is a constant. After the same procedure has been performed
 for all the six pairs, it is finally found that
\begin{eqnarray}
\bar{s}_3 =
\sigma (\phi ,s_a ,\bar{s}_1 ,\bar{s}_2 )\propto 
\phi^{\xi_\phi}
\prod^3_{a=1} (s_a)^{\xi_a} \prod^2_{b=1} (\bar{s}_b)^{\bar{\xi}_b},
\label{0024}
\end{eqnarray}
where the exponents $\xi_\phi$,$\xi_a$ and $\bar{\xi}_b$ are constants.
 Solving eqn (\ref{0024}) for $\phi$ leads to 
 the following result.
\begin{eqnarray}
\phi=C_o \prod^3_{a=1} (s_a)^{-\nu_a} \prod^3_{b=1}
(\bar{s}_b)^{-\bar{\nu}_b}
=C_o \Delta ,
\label{0025}
\end{eqnarray}
 where $C_o$, $\nu_a$ and $\bar{\nu}_b$ are also constants.
 By introducing a new function $F$ as
\begin{eqnarray}
F(x; r_A ,r_B)
=
\Omega
(C_o x,C_o x,C_o x,C_o x,C_o x,C_o x ;r_A ,r_B),
\label{0026}
\end{eqnarray}
the final result (\ref{13}) is achieved using
 eqns (\ref{0015}) and (\ref{0025}) .

Next let us prove that the Tsallis-type bi-composable entropy
takes the form (\ref{14}). Firstly let us impose the
 grand pseudo additivity (\ref{5}) on the bi-composable 
 entropy (\ref{13}). Then it turns out that the function $F$
  possesses the following property:
\begin{eqnarray}
F(xy ;r_A ,r_B)
&=&F(x ;r_A ,r_B)+F(y ;r_A ,r_B)\nonumber\\
&&+(1-Q)F(x;r_A ,r_B)
F(y; r_A ,r_B)
\label{0027}
\end{eqnarray}
 for arbitrary parameters $x$ and $y$. Differentiating eqn (\ref{0027})
  with respect to $y$ and taking $y=1$ yield
\begin{eqnarray}
x\frac{\partial F}{\partial x}
=F'(1 ;r_A ,r_B) +(1-Q)F'(1;r_A ,r_B)F(x;r_A ,r_B),
\label{0028}
\end{eqnarray}
where $F'(1;r_A ,r_B)$ is derivative of $F$ at $x=1$. Eqn (\ref{0028})
 can be integrated and we get the following solution:
\begin{eqnarray}
F(x ;r_A ,r_B)=-\frac{1}{1-Q}+C' x^\eta
\label{0029}
\end{eqnarray}
where $C'$ and $\eta$ are constants. Substituting eqn (\ref{0029})
 into eqn (\ref{0027}) fix the constant $C'$ as
\begin{eqnarray}
C' =\frac{1}{1-Q}
\end{eqnarray}
and finally the function $F$ is determined as
\begin{eqnarray}
F(x ;r_A ,r_B )=-\frac{1}{1-Q}
\left[1-x^\eta \right].
\label{0030}
\end{eqnarray}
Therefore, redefining as
\begin{eqnarray}
&&
\eta \nu_a \rightarrow \nu_a,
\\
&&
\eta \bar{\nu}_b \rightarrow \bar{\nu}_b 
\end{eqnarray}
 in the definition of $\Delta$, it is concluded that the most general form  
 of the Tsallis-type bi-composable entropy is given by eqn (\ref{14}).

\section{A Regular Example of Tsallis-Type Bi-Composable Toy Model}
Let us discuss  a plain case that 
 the grand index $Q$ and the exponents $\nu_a$, $\bar{\nu}_b$ 
 behave as functions only 
 in terms of  $r_A$ and $r_B$. 
 Then in the functions 
 the permutation symmetry between the subsystems
 $A$ and $B$ should be preserved:
\begin{eqnarray}
&&
Q(r_A ,r_B)=Q(r_B ,r_A),
\\
&&
\bar{\nu}_b (r_A ,r_B) =\nu_b (r_B ,r_A),\ \ (b=1,2,3).
\end{eqnarray}
Furthermore when we take $r_A =r_B$, that is, $q_A =q_B$,
 the composite entropy must be reduced into the original Tsallis entropy
 (\ref{39}). Hence we require the following boundary conditions:
\begin{eqnarray}
&&
Q(r,r)=\frac{1}{r} =q,\label{40}
\\
&&
\nu_1 (r,r) =\frac{1}{2r}=\frac{q}{2} ,\label{41}
\\
&&
\nu_2 (r,r)=\nu_3 (r,r)=0.\label{42}
\end{eqnarray}
Then let us introduce a entorpy form as
\begin{eqnarray}
\hat{S}_{A+B}
&=&
-\frac{r_A (r_A -1)^2 +r_B (r_B -1)^2}
{(r_A -1)(r_B-1) (r_A +r_B -2) }
\nonumber\\
&&\times
\left[
1-
\left(
X_1^{r_B -1} \bar{X}_1^{r_A -1} 
\left[
\frac{X_2}{\bar{X}_2}
\right]^{r_A -r_B}
\right)^{
\frac{2-r_A -r_B }
{2\left[ 
r_A (r_A -1)^2 +r_B (r_B -1)^2
\right]}
}
\right].\label{46}
\end{eqnarray}
Clearly if $r_A =r_B$ is taken, the entropy in eqn (\ref{46})
is reduced into the original Tsallis entropy.
 Also it can be pointed out that 
 eqns (\ref{36}), (\ref{37}) and (\ref{38}) certainly hold in eqn (\ref{46}),
 thus the entropy  satisfies the H-theorem for near equilibrium
and factorized states.  

Here one might worry about a "singularity" at 
\begin{eqnarray}
r_A +r_B =2
\end{eqnarray}
in eqn (\ref{46}). But fortunately
it is explicitly checked that 
if one takes $r_B \rightarrow 2-r_A$ limit, 
any singularity does not take place
 and the entropy really 
 has well-defined finite form as
\begin{eqnarray}
\lim_{r_B \rightarrow 2-r_A }\hat{S}_{A+B}
=
-\frac{1}{2(1-r_A)}
\ln
\left(
\frac{\bar{X}_1}{X_1}
\left(
\frac{X_2}{\bar{X}_2}
\right)^2
\right).\label{001}
\end{eqnarray}

The limit of $r_A \rightarrow 1$ is also obtained as a regular form. 
\begin{eqnarray}
\lim_{r_A \rightarrow 1} \hat{S}_{A+B}
=
-\frac{X_{1;2}}{2X_2} -\frac{1}{2}\bar{X}_{2;I}
+\frac{1}{2(1-r_B  )} \ln X_2 +\frac{1}{2(1-r_B)}\ln \bar{X}_3,\label{004}
\end{eqnarray}
where
\begin{eqnarray}
X_{1;2}&=&
\lim_{r_A \rightarrow 1}
\frac{X_1 -X_2}{r_A -1}
\nonumber\\
&=&
-r_B \sum^M_{j=1} \left(\sum^N_{i=1} P_{ij} \right)^{r_B}
 \ln \left(\sum^N_{i'=1}P_{i' j} \right)
 +r_B \sum^M_{j=1}
 \left(\sum^N_{i=1}P_{ij} \right)^{r_B -1}
 \left(\sum^N_{i' =1} P_{i' j} \ln P_{i'j} \right),
\nonumber\\
\bar{X}_{2;I} &=&
\lim_{r_A \rightarrow 1}
\frac{\bar{X}_2 -1}{r_A -1}
\nonumber\\\
&=&
\sum^N_{i=1}
\left(\sum^M_{j=1} P_{ij} \right)
\ln \left( 
\sum^M_{j'=1} P_{ij'}
\right).\nonumber
\end{eqnarray}

\end{document}